\documentclass[preprint]{aastex}
\usepackage{graphicx}
\usepackage{epsfig}
\usepackage{amsmath}
\usepackage{lineno}
\usepackage{comment}

\begin{document}
\shorttitle{}
\shortauthors{Fang $\&$ Fan}
\title{$\delta$-Sunspot Formation in Simulation of Active-Region-Scale Flux Emergence}

\author{Fang Fang, Yuhong Fan}
\affil{High Altitude Observatory, National Center for Atmospheric Research, 
3090 Center Green Dr, Boulder, CO 80301}


\begin{abstract}
$\delta$-sunspots, with highly complex magnetic structures, are very productive
in energetic eruptive events, such as X-class flares and homologous eruptions. 
We here study the formation of such complex magnetic structures 
by numerical simulations of magnetic flux emergence from the convection zone
into the corona in an active-region-scale domain. 
In our simulation, two pairs of bipolar sunspots form on the
surface, originating from two buoyant segments of a single subsurface 
twisted flux rope, following the approach of \cite{toriumi2014}. 
Expansion and rotation of the emerging fields in the two bipoles drive the two 
opposite polarities into each other with apparent rotating motion,
producing a compact $\delta$-sunspot with a sharp polarity inversion line. 
The formation of the $\delta$-sunspot in such a realistic-scale domain produces 
emerging patterns similar to those formed in observations, 
e.g. the inverted polarity against 
Hale's law, the curvilinear motion of the spot, strong transverse field 
with highly sheared magnetic and velocity fields at the PIL. 
Strong current builds up at the PIL, giving rise to reconnection, which produces 
a complex coronal magnetic connectivity with non-potential fields 
in the $\delta$-spot overlaid by more relaxed fields connecting the two polarities 
at the two ends. 
\end{abstract}


\section{Introduction} \label{intro}

$\delta$-sunspots, where two spots of opposite polarities share the same penumbrae,
are of particular interest due to their high productivity of powerful flares and 
coronal mass ejections (CMEs) 
\citep{hagyard1984,tanaka1991,schmieder1994,sammis2000,liu2001}. 
Based on their characteristics during the formation observed at Big Bear Solar 
Observatory, \cite{zirin1987} categorize 
the $\delta$-sunspots into three types: island $\delta$ which emerge at once but 
with different dipoles intertwined, single $\delta$ with large satellite dipoles 
emerging in the penumbra of a large spot, and a third and most popular type formed 
by collision 
of two bipolar spot groups. Regardless of their types, all the $\delta$ spots, 
which form from opposite polarities of different dipoles and die together, 
are short-lived complex structures with highly sheared field lines at the PIL.  
They manifests their complexity on the photosphere as highly structured active 
regions, with non-potential coronal fields in the 
form of sheared and stretched field lines connecting the sunspots 
\citep{patty1986,tanaka1991, lites1995, leka1997}, 
storing a large amount of free magnetic energy available for the 
flare activities. 
In many occasions, the footpoints of the $\delta$-spots exhibit shearing motion 
at the PIL and rotation of the spots \citep{tanaka1991, leka1996}, 
which further stretch and twist the magnetic field lines. 
The combination of non-potential magnetic field and shearing, rotating motion 
at the PIL produces a magnetic configuration that is prone to 
magnetic reconnection and eruption, which might be repetitive in the presence 
of energizing motions \citep{fang2012b}. 

According to the observed features of compact and complex $\delta$-spots regions, 
\cite{zirin1987} point out that they may be the result of a 
subsurface phenomenon which locks the two spots closely together and it is 
therefore important to investigate the subsurface processes that form and maintain these 
complex spots and build up the free energy. 
Later \cite{tanaka1991} and \cite{leka1996} 
suggest that $\delta$-sunspots can be formed by the emergence of highly twisted 
magnetic flux tubes from the convection zone, which can explain self-consistently
the shearing and rotation motion as well as the compact pattern of opposite 
polarities. 
This concept then motivated series of numerical simulations 
on helical kink instability of highly twisted rising flux tubes emerging from 
the interior \citep{linton1998, linton1999,fan1999}. 
The kinked evolution of the rising tubes produces 
a compact magnetic bipolar region, with polarity order reversed from 
Hale's law, with great similarities to observed features of $\delta$-sunspots. 
The braided kinked magnetic structure keeps the two opposite polarities closely 
together and drives the rotation around each other. 
\cite{tanaka1991} also proposed a model which involved the successive emergences
of two twisted magnetic flux ropes which may or may not be interconnected under the surface, 
to explain the evolution of complex $\delta$-spots. 
Recent study on the quadrupolar active region 11158 with colliding $\delta$-spots 
\citep{sun2012} found that the magnetic fields were highly sheared at the PIL 
carrying strong current, and significant free energy was injected into the atmosphere.
Later \cite{toriumi2014} 
considered two scenarios of the subsurface emerging structures for simulating the formation 
of $\delta$-spots: two flux ropes each with one buoyant section, and 
a single flux rope with two buoyant sections. 
They found that only the emergence of the two buoyant sections from a single flux rope 
forms a sharp PIL in the domain. 
Following this work, here we also carry out a simulation 
of the emergence of a twisted flux rope with two buoyantly rising sections, 
but in a domain of size comparable to active regions to 
study whether compressed $\delta$-spot like structures of realistic active region 
size scales can still form. 
In the following sections, we describe the numerical setup
of our simulation in Section \ref{method}, and investigate the formation of $\delta$-sunpots 
and the photospheric flows and energy flux in Section \ref{formation} and \ref{flows}, respectively,
followed by conclusions in Section \ref{conclusion}.

\section{Simulation Setup} \label{method}

\begin{figure*}[ht!]
  \begin{minipage}[t] {1.0\linewidth}
    \begin{center}
      \includegraphics[width=100mm]{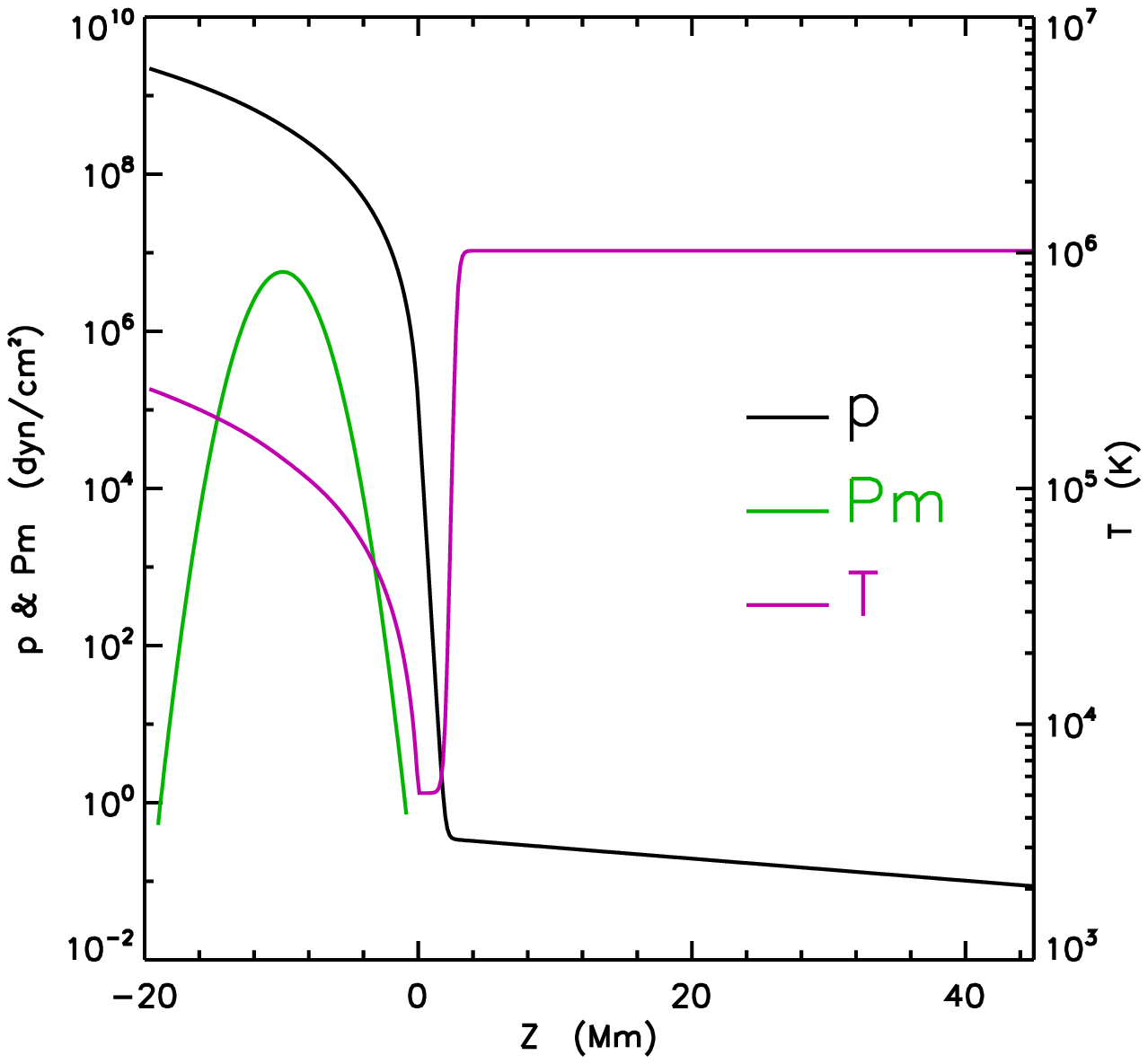}
    \end{center}
  \end{minipage}\hfill
  \caption{Initial vertical stratification of pressure (black) and temperature 
    (purple) in the domain. The green line shows the distribution of 
    magnetic pressure in a cross-section cut of the flux rope.}
  \label{initatm}
\end{figure*}

\begin{figure*}[ht!]
  \begin{minipage}[t] {1.0\linewidth}
    \begin{center}
      \includegraphics[width=80mm]{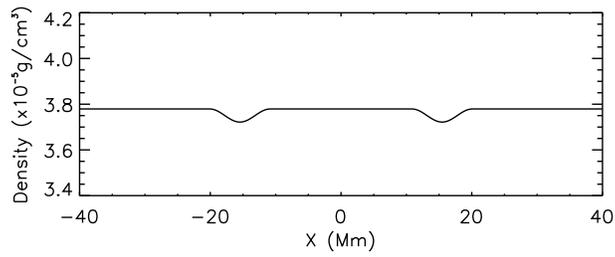}
    \end{center}
  \end{minipage}\hfill
  \caption{Distribution of the initial density along the axis of the flux rope with 
    two buoyant sections centered at x = $\pm$15 Mm.}
  \label{initrho}
\end{figure*}

We solve the ideal MHD equations within the Block Adaptive Tree 
Solar-wind Roe Upwind 
Scheme (BATS-R-US) that was developed at the University of Michigan 
\citep{powell1999,toth2012}:
\begin{equation}
  \label{mass}
  \frac{ \partial \rho}{\partial t} + \nabla \cdot (\rho {\bf u}) = 0, 
\end{equation}
\begin{equation}
  \label{momentum}
  \frac{\partial(\rho{\bf u})}{\partial t}+\nabla \cdot \left[\rho{\bf u}  
    {\bf u}+\left(p+\frac{{\bf B}{\bf B}}{8\pi}\right)\mathbf{I}-\frac{{\bf
        B}{\bf B}}{4\pi}\right]
  = \rho{\bf g},
\end{equation}
\begin{equation}
  \label{energy}
  \frac{\partial E}{\partial t} + \nabla \cdot \left[\left(                
    E + p + \frac{{\bf B}\cdot{\bf B}}{8\pi}\right){\bf u} - \frac{
      ({\bf u}\cdot{\bf B}){\bf B}}{4\pi}\right]
  = 0,
\end{equation}
\begin{equation}
  \label{induction}
  \frac{\partial {\bf B} }{\partial t} = \nabla\times ({\bf u}\times{\bf B} ),
\end{equation} 
where $\rho$, ${\bf u}$, $E$, $p$, $T$, ${\bf B}$ and ${\bf g}$ are the density, 
velocity, total energy density, pressure, temperature, magnetic field and 
gravitational acceleration, respectively. 
The domain of our simulation covers 158(X)$\times$138(Y)$\times$79(Z) Mm$^3$, 
extending 20~Mm below the photosphere and 59~Mm into the corona with 
an initial background stratification of atmosphere in hydrostatic equilibrium
\citep{fan2001}. 
The cell sizes range from 38.6 to 618.68 km in the vertical direction 
from the interior to coronal height and 77.3 to 1237.4 km in 
horizontal directions. 
A horizontal flux rope in the {\bf x} direction is inserted at depth Z = -10~Mm, 
with the magnetic field given by ${\bf B} = B_{0}e^{-r^{2}/a^{2}}\hat{\boldsymbol x} + 
qrB_{0}e^{-r^{2}/a^{2}}\hat{\boldsymbol\theta}$, where $B_{0} = - 12$~kG, 
is the strength of the magnetic fields at the central axis of the rope, 
$\hat{\boldsymbol\theta}$ is the azimuthal direction in the cross section of the flux rope, 
$q = 0.50/a$, is the rate of twist per unit length, $a = 3.0$~Mm is the radius of 
Gaussian decay, and $r$ is the distance from the axis of the flux rope. 
Fig.~\ref{initatm} shows the vertical stratification of 
pressure and temperature in the domain. The green line represents the distribution
of magnetic energy inside the flux rope in a cross-section cut. The plasma beta 
at the central axis of the flux rope, $\beta = 66$. 
The initial rope is density-depleted at two segments centered at $x = \pm$15~Mm 
as shown in Fig.~\ref{initrho}, 
with perturbation as:
\begin{equation}
\Delta\rho = - \frac{0.5[B_0e^{(-r^2/a^2)}]^2}{p}\rho_0\begin{cases}
  -\eta + \frac{(1+\eta)\cos[\pi(x - x_+)/\lambda]}{2}, & \text{if}~|x-x_+| \leq \lambda \\
  -\eta + \frac{(1+\eta)\cos[\pi(x - x_-)/\lambda]}{2}, & \text{if}~|x-x_-| \leq \lambda \\
  -\eta,     & \text{otherwise.}
\end{cases}
\end{equation}
Here $\rho_0$ is the initial density value in hydrostatic equilibrium, 
buoyancy factor $\eta$ = 0.15, 
and x$_\pm$ = $\pm$ 15 Mm, $\lambda$ = 4.6 Mm is the half length of each buoyant segments. 
And the thermal pressure inside the flux rope is modified to maintain 
the total pressure balance. 

\section{Formation of $\delta-$sunspots} \label{formation}

\begin{figure*}[ht!]
  \begin{minipage}[t] {1.0\linewidth}
    \begin{center}
      \includegraphics[width=100mm]{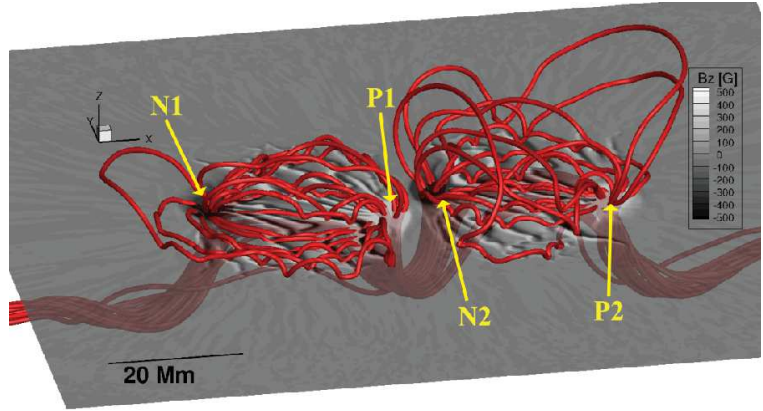}
    \end{center}
  \end{minipage}\hfill
  \caption{3D structure of the M-shaped magnetic field lines (red lines) 
    in the domain at time t = 2.5 hrs. 
    The plane shows the photospheric magnetogram. }
  \label{3dt0230}
\end{figure*}

The density-depleted sections, centered at x = $\pm$ 15 Mm of the initial flux rope, 
rise in the convection zone under buoyancy.  
The dramatic decrease of pressure in the ambient plasma with altitude gives rise 
to the vast expansion in the two rising sections 
of the flux rope as it emerges. Fig.~\ref{3dt0230} illustrates the three-dimensional
(3D) structure of the magnetic flux rope in the simulation domain after 2.5 hours 
of evolution with the gray plane showing photospheric vertical magnetic field $B_z$. 
Each of the two buoyant section rises and expands, bending the initial flux rope and 
forming an M-shaped configuration of magnetic fields 
extending from the convection zone into the corona. 
At the same time, the mass draining in the buoyant sections
down along the field lines under gravitational force 
further enhances the buoyancy in the rising 
sections. Gravitational force of the heavy material 
keeps the central segment between the two rising sections down in the convection zone, 
while the two buoyant sections emerge. 
The downward push from the drained mass at the central segment 
between the rising bipoles maintains 
the configuration of multiplicity of bipoles in the domain, and prevents the multiple 
bipoles from merging into one pair of bipoles. 
Note that in the real solar convection zone, the downward convective flow may make an 
additional contribution to the downward push and facilitate the formation of multiple bipoles. 
The emerged sections expand dramatically due to the stratification of density in the 
surrounding plasma and especially when it approaches the photosphere, density drops by 
6 orders of magnitude over a thin layer of 2 Mm, driving an enormous expansion 
of the flux rope in the horizontal direction. This expansion gives rise to 
an unstable situation where high-density material is supported by low-density 
plasma at the top of the flux rope. Under this condition, Rayleigh-Taylor instability 
takes place at the top boundary of the emerging fields 
and produces the undular structure on the field lines 
as represented by the red lines in Fig.~\ref{3dt0230}. 
The grey plane of photospheric $B_z$ also shows strands of opposite polarities in the center 
of each bipole, manifesting the effect of Rayleigh-Taylor instability on the photospheric plane. 
\cite{isobe2005, isobe2006} have shown that magnetic Rayleigh-Taylor instability 
in simulations of flux emergence into preexisting coronal field also takes place in the top-heavy 
configuration caused by the expansion of the emerging dome and the draining of plasma along the 
field lines, and produces dense filaments mimicing H$\alpha$ arch filaments, 
which may cause intermittent heating and the hot and cold loops observed in extreme 
ultraviolet images \citep{yoshimura1999}.

\begin{figure*}[ht!]
  \begin{minipage}[t] {1.0\linewidth}
    \begin{center}
      \includegraphics[width=100mm]{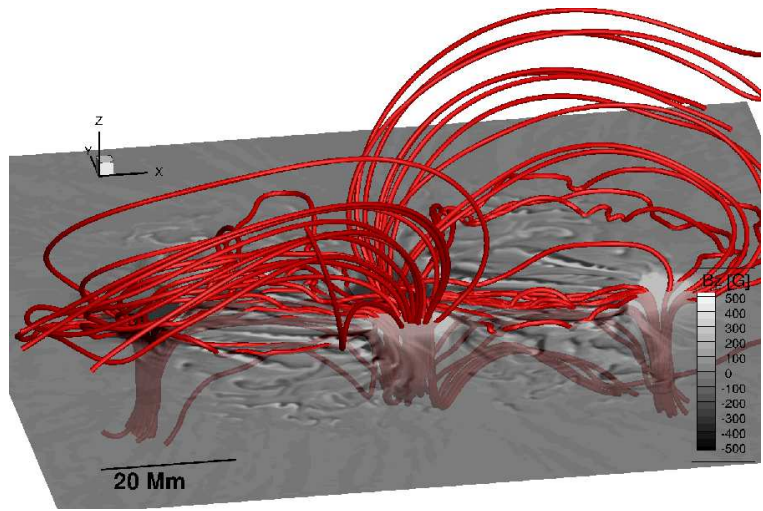}
    \end{center}
  \end{minipage}\hfill
  \caption{Same as Fig.~\ref{3dt0230} but at time t = 3.5 hrs.}
  \label{3dt0330}
\end{figure*}

\begin{figure*}[ht!]
  \begin{minipage}[t] {1.0\linewidth}
    \begin{center}
      \includegraphics[width=100mm]{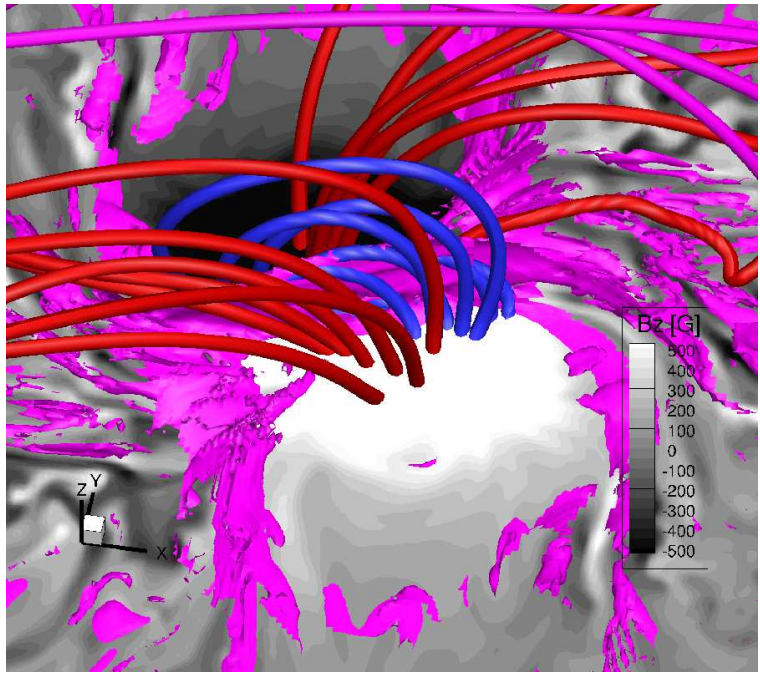}
    \end{center}
  \end{minipage}\hfill
  \caption{The $\delta$-spot region at time t = 4.5 hrs with magenta isosurfaces of 
    $|J|/|B|$ = 100 in regions of B$_h >$ 300 G and colored lines representing 
    the magnetic field lines.}
  \label{t0430j}
\end{figure*}

\begin{figure*}[ht!]
  \begin{minipage}[t] {1.0\linewidth}
    \begin{center}
      \includegraphics[width=100mm]{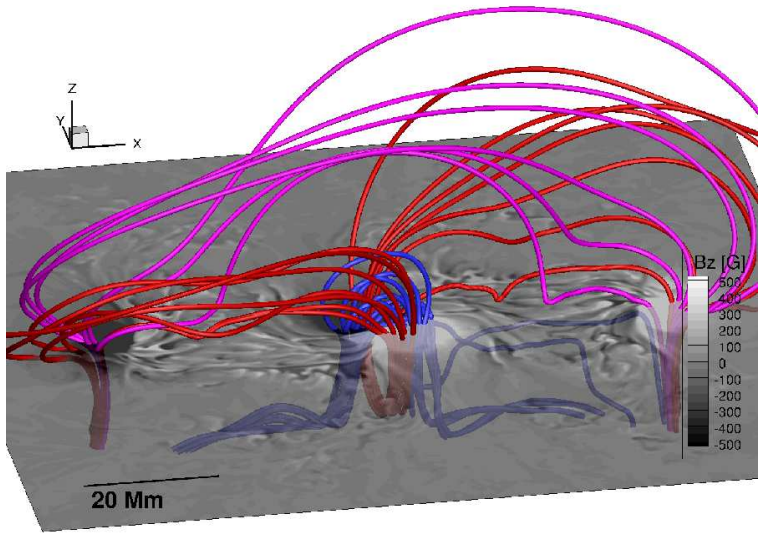}
    \end{center}
  \end{minipage}\hfill
  \caption{Same as Fig.~\ref{3dt0230} but at time t = 4.5 hrs. 
  The magenta and blue lines are showing the overlying and low-lying magnetic field 
  lines after the reconnection, respectively. 
  The red lines show the field lines 
  connecting the four polarities.}
  \label{3dt0430}
\end{figure*}

The initial flux rope has a right-hand twist with q = 0.5, which causes 
the bipoles to initially emerge with its polarity inversion line at an angle to 
the axis of the flux rope. 
As the expansion of the buoyant emerging bipoles continues, the polarities in 
each bipole start to shift toward the central axis of the initial flux rope. 
At the center of the domain, the two polarities
next to the dip in the M-shaped structure are pushed together by the expansion 
in each of the emerging bipoles, as well as the shifting motion toward the 
central axis, as shown in Fig.~\ref{3dt0330}. These motions drive the two opposite 
polarities together to 
naturally form a sharp PIL between the the two polarities P1 and N2 (shown by 
arrows on Fig.~\ref{3dt0230}) from the two 
bipoles. 
Note that underneath the photosphere, the two polarities P1 and N2 at the PIL are 
connected by a subsurface U-loop structure, while for each of the bipoles, 
i.e. N1-P1 and N2-P2, the opposite polarities are connected by $\Omega$-loops 
above the surface. 
At time t = 3.5 hrs, the magnetic 
fields, shown by the red lines in Fig.~\ref{3dt0330}, 
expand further into the corona, forming overlying coronal loops 
connecting each of the bipoles. Importantly, along the PIL in the center, 
the magnetic fields from the two opposite polarities P1 and N2 are anti-parallel 
to each other, while their footpoints are pushed together during the emergence 
and expansion. 
In such a configuration of anti-parallel magnetic fields, 
strong current sheet builds up at the PIL where the two polarities P1 and N2 are moving 
toward each other, as shown by the magenta surfaces in Fig.~\ref{t0430j} 
between the red lines connecting N1-P1 and N2-P2. 
With the continuing convergence of the two polarities during the 
expansion of the two emerging bipoles, current keeps thinning until 
reconnection takes places at the PIL, between the two nearby polarities P1 and N2.
The three groups of lines in Fig.~\ref{3dt0430} illustrate the complex coronal 
magnetic fields in our simulation domain. 
The reconnection forms the large-scale overlying loops connecting the two polarities
at the two ends of the flux rope N1 and P2, as shown by the magenta lines in 
Fig.~\ref{3dt0430}, as well as a set of small loops connecting the nearby 
polarities P1 and N2 in the center shown by the blue lines. The connectivity 
of the initial magnetic fields is colored by red, which shows an M-shaped structure. 
By comparing Fig.~\ref{3dt0330} and \ref{3dt0430}, we notice the change of the 
magnetic connectivity under the continuous expansion and emergence. 
We note that it is the reconnection at the PIL that produces the complexity 
in the structure of magnetic field. 
At this time, $\delta$-spot forms in the center with complex
magnetic connectivity among the polarities and a compact magnetic 
structure at the PIL (see Fig.~\ref{t0430j}). 

\begin{figure*}[ht!]
  \begin{minipage}[t] {1.0\linewidth}
    \begin{center}
      \includegraphics[width=80mm]{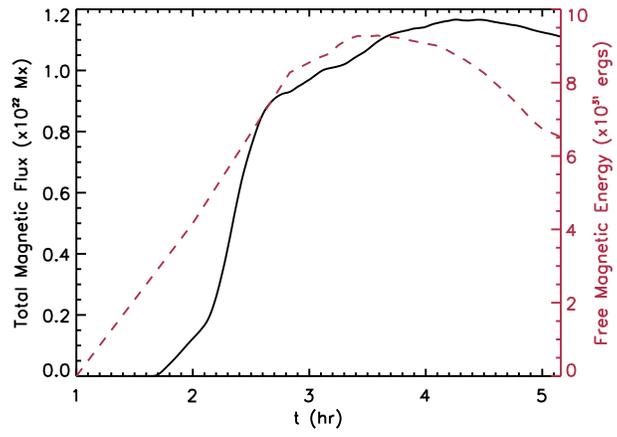}
    \end{center}
  \end{minipage}\hfill
  \caption{Evolution of total unsigned magnetic flux at the photosphere 
    in the computational domain and the coronal free magnetic energy.}
  \label{totflux}
\end{figure*}

The evolution of total unsigned flux on the photosphere 
is shown in Fig.~\ref{totflux}. We observe a fast growth
in the total flux until about time t = 2.6 hrs, when the two $\Omega$-shaped 
structures come into contact and collide. Until this time, the expansion and emergence 
of each structure are independent of each other and the magnetic flux simply 
grows into regions of less magnetic pressure, filling the space between the 
two buoyant segments rapidly. After time t = 2.6 hrs, 
the majority of the axial flux reaches the photosphere, 
as shown in Fig.~\ref{3dt0230}, forming a quadupolar structure on the surface. 
After that, the growth of the total unsigned flux is slowed down, indicated 
by the less steep gradient from time t = 2.6 to 4.5 hrs, when 
the two polarities P1 and N2 collide into each other. 
During the collision, strong current builds up at the sharp PIL, 
and reconnection takes place between the two polarities P1 and N2, 
producing flux cancellation at the PIL. We evaluate the potential magnetic energy 
from the photospheric magnetogram and calculate total free magnetic energy in 
the corona by subtracting the potential energy from the magnetic energy in the 
coronal fields, as shown by the red dashed line in Fig. \ref{totflux}. 
The free energy keeps increasing with the flux emergence until 3.6 hrs, when 
the reconnection at the PIL starts to release the stored free magnetic energy 
by forming shorter loops shown by the blue lines in Fig. \ref{3dt0430}
with a total drop of 2.8$\times$10$^{31}$ ergs in free energy. The release of 
free energy is comparable with the estimated value of 3.4$\times$10$^{31}$ ergs in 
AR 11158 as in \cite{sun2012} although the energy release is more gradual in 
the simulation with no eruptions.


\section{Photospheric Flows and Energy Flux} \label{flows}

\begin{figure*}[ht!]
  \begin{minipage}[t] {1.0\linewidth}
    \begin{center}
      \includegraphics[width=120mm]{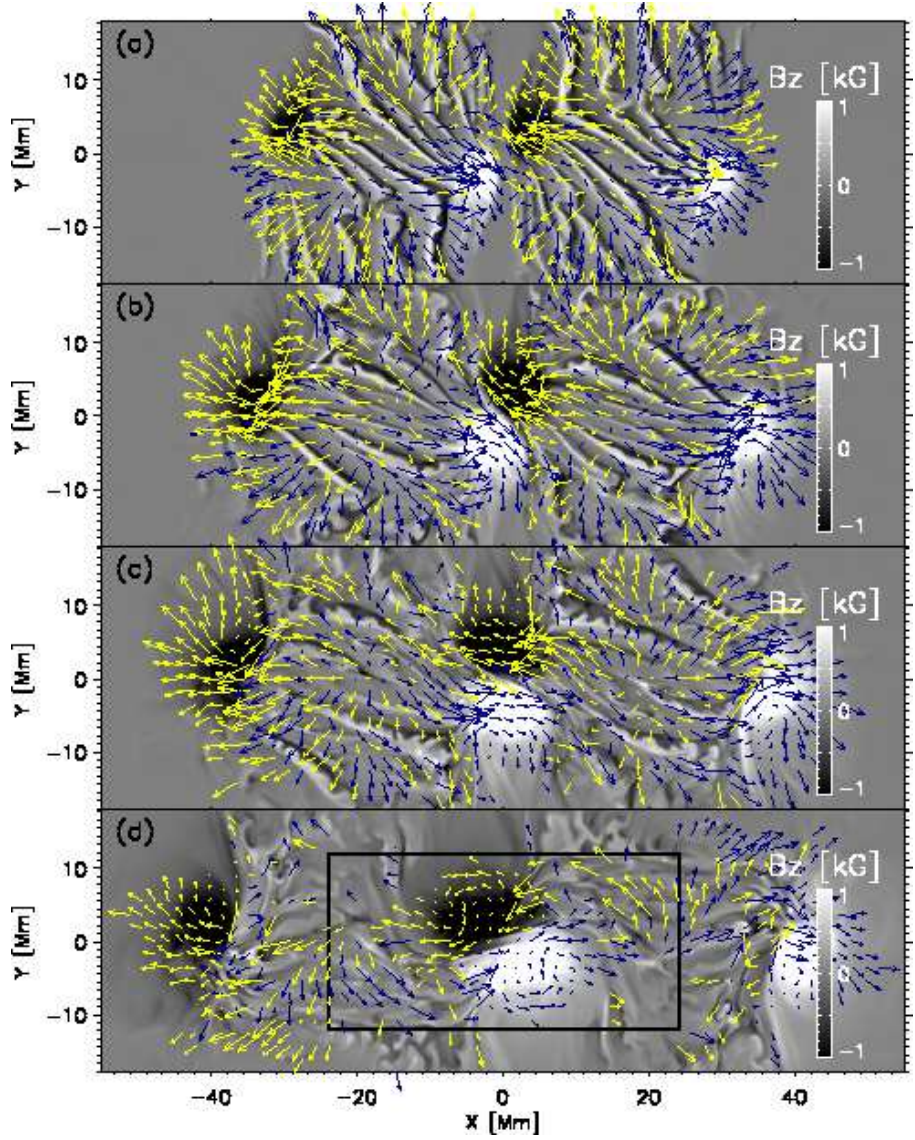}
    \end{center}
  \end{minipage}\hfill
  \caption{Magnetogram at time t = 2.5 (a), 3.0 (b), 3.5 (c) and 4.5 (d) hrs. 
    Blue and yellow arrows show the horizontal velocity in positive and negative 
    polarities with $|B_z| > $ 800 G, respectively. 
    A movie showing the evolution of
    magnetogram with horizontal velocities is available online.}
  \label{bzuxy}
\end{figure*}

At the photosphere, the two buoyant sections of magnetic flux rope centered at 
x = $\pm$ 15 Mm 
first emerge as two separate bipoles at time t = 2.5 hrs, 
as shown in Panel (a) of Fig.~\ref{bzuxy}. 
The distinct emergence of the two sections can also be viewed in 
the 3D magnetic structure shown at the same time in Fig.~\ref{3dt0230}. 
Until then, the emergence of the two $\Omega$-shaped structures remains independant of each other, 
with only their footpoints being connected in the convection zone. 
The blue and yellow arrows represent the horizontal velocity in the positive and negative 
polarities, respectively. 
It is clear that at this time the velocity fields in N1-P1 and N2-P2 are mostly 
pointing outward from the center of each emerging section, 
representing the expansion of the emerging domes in a stratified atmosphere with 
dramatic drop in the pressure of the ambient plasma. 
Due to the initial seperation between the two density-depleted sections and the 
mass draining into the central dip, 
the two polarities P1 and N2 in the center remain separated at this time t = 2.5 hrs and 
unaffected by each other with a distance from each other as shown in Fig.~\ref{3dt0230}.
The continuous expansion during 
the emergence in the stratified atmosphere drives the outward motion of each polarities 
from the center of 
each bipole, as shown by the arrows representing horizontal velocity. 
Inevitably, the two $\Omega$-shaped emerging domes come into contact with each 
other and the two opposite polarities P1 and N2 are pushed together during 
the expansion, as shown by the velocity arrows in Panel (b) of Fig.~\ref{bzuxy}.
At the same time, a sharp PIL with high field gradient forms between P1 and N2 
as they are pushed and squeezed into each other, 
which is found to be the preferential regions of flares and often used 
as an observational predictor of solar eruptive activities 
\citep{leka2003a, schrijver2007, falconer2008, mason2010}. 
More importantly, as the two polarities cannot annihilate with each other 
during the build up of sharp magnetic gradient, it is necessary that 
the two spots P1 and N2 take another curvilinear path around each other 
during the expansion of the two bipoles, 
which is also found in the observations of $\delta$-spots formed by 
collision of non-paired spots \citep{tang1983}. 
The horizontal velocity in the positive (P1) and negative (N2) 
polarities, as shown by the blue and 
yellow arrows, respectively, runs almost anti-parallel to each other along the PIL, 
exhibiting a strong shearing pattern between the two polarities. 
The velocity shear, driven by the spot motion \citep{zirin1987} -- flux emergence 
in our case, persists
during the interaction of the two polarities P1 and N2, as shown by the 
anti-parallel pattern of velocity arrows in Panel (b)-(d) at the PIL. 
It streches the two sets of magnetic field lines connecting N1-P1 and N2-P2 respectively 
at the PIL in opposite directions and forms a set of highly sheared field lines 
at the interface between P1 and N2, as shown in Fig.~\ref{3dt0330}.
As discussed in Section \ref{formation}, reconnection takes place at the PIL 
and forms a set of complex coronal structures as shown in Fig.~\ref{3dt0430}. 
The reconnection here, similar to that in \cite{vanballegooijen2007} 
where two neigbouring $\Omega$-loops interact and reconnect at the PIL, 
releases the free magnetic energy by reconfiguring the stretched fields at the PIL into 
more relaxed ones.
The continuous shearing motion represented by the arrows adjacent to the PIL both in Panel (d) 
of Fig.~\ref{bzuxy} and the movie available online then moves the footpoints of the 
post-reconnection low-lying field lines above the PIL in the opposite directions 
and stretch them into elongated and sheared magnetic structures. 
The highly sheared magnetic configuration at the PIL is believed to be one of 
the preceding signatures of great flares in $\delta$-spots \citep{zirin1987}, 
during which the free energy is released 
by relaxing and shortening the magnetic field lines. 

\cite{tang1982} also pointed out that the the polarity axis in $\delta$-spots 
often deviates from Hale's law, based on sunspot observations from Mount Wilson. 
In our simulation, as shown by the evolution of polarities at the photopshere in 
Panel (a)-(d) of Fig.~\ref{bzuxy}, 
we naturally get the inverted polarity of the $\delta$-spot P1 and N2 
with the current configuration of the subsurface flux rope from the 
collision of the two emerging bipoles, 
as the following polarity of the $\delta$-spot is the leading
one in the following bipole. 
During its evolution, however, the rotation motion of the $\delta$-spot 
drives the two polarities P1 and N2 in a direction towards obeying the Hale'slaw. 
The counterclockwise rotation 
of the $\delta$-spot is one of the manifestations of the curvilinear paths of the two 
polarities P1 and N2, which is ultimately driven by the expansion in each bipole. 

\begin{figure*}[ht!]
  \begin{minipage}[t] {1.0\linewidth}
    \begin{center}
      \includegraphics[width=100mm]{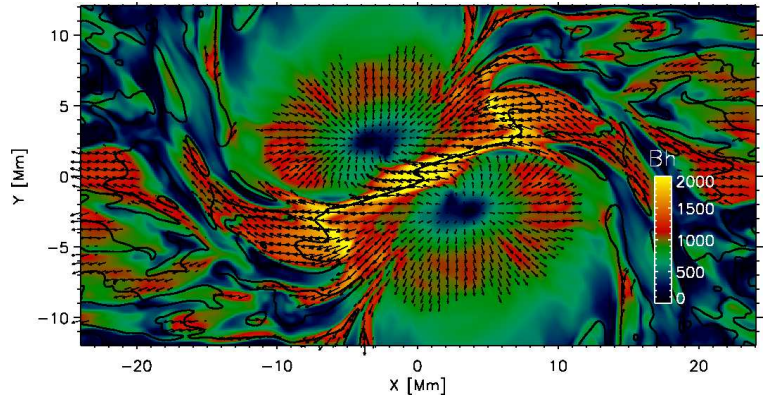}
    \end{center}
  \end{minipage}\hfill
  \caption{The horizontal magnetic field with color showing strength and arrows showing directions 
    in the area outlined by the rectangle 
    in Panel (d) of Fig.~\ref{bzuxy} at time t = 4.5 hrs. The black lines represent the PILs.}
  \label{zoombh}
\end{figure*}

As a result of the cuvilinear path of the spots, the magnetic fields between the two 
spots take the form of highly compressed and streched field lines at the PIL. 
The blue lines in Fig.~\ref{3dt0430} show a group of sheared reconnected fields 
among the more relaxed field lines connecting all four polarities in the domain shown by 
the red lines. Fig.~\ref{zoombh} shows the strength of the horizontal magnetic field, 
i.e. $B_h = (B_x^2+B_y^2)^{1/2}$, 
in the $\delta$-spot region outlined by black rectangle in Fig.~\ref{bzuxy}. 
Approaching the PIL, the horizontal field strength 
quickly increases to above 2 kG, 
as compared to the nearly vertical field in the center of each spot 
shown by the dark blue color. 
These strong transverse magnetic fields with strength comparable to that in the
umbrae are confined to long, extremely narrow channels in $\delta$-spots in Fig.~\ref{zoombh}
and defined as ``magnetic channels" by \cite{zirin1993a}.
In addition, the arrows in Fig.~\ref{zoombh} representing the horizontal magnetic fields
align almost parallel to the PIL, indicating that the 
transverse fields of kG-strength are along the PIL and thus highly non-potential \citep{low1982}, 
with the free magnetic energy stored which could possibly be released in eruptions.

\begin{figure*}[ht!]
  \begin{minipage}[t] {1.0\linewidth}
    \begin{center}
      \includegraphics[width=100mm]{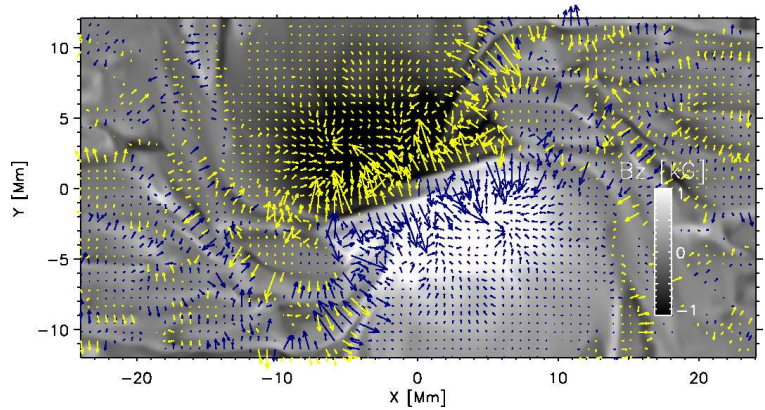}
    \end{center}
  \end{minipage}\hfill
  \caption{Vertical magnetic field B$_z$ in the area outlined by the rectangle 
    in Panel (d) of Fig.~\ref{bzuxy}
    at time t = 4.5 hrs with blue and yellow arrows showing 
    the horizontal Lorentz force in positive and negative polarities, respectively.} 
  \label{lorentz}
\end{figure*}

Besides the shearing motion that forms highly elongated and sheared field lines 
adjacent to the PIL,
a converging motion occurs in the outer periphery of the encountering spots, 
and keeps pushing the two spots P1 and N2 together. 
This converging motion results from the expansion and squeeze from the emerging 
bipoles of the writhed emerging rope, 
as shown by the 3D magnetic structure in Fig.~\ref{3dt0330}. 
In Fig.~\ref{lorentz}, we calculated the Lorentz force in the $\delta$-spot 
region outlined by the rectangle in Panel (d) of Fig.~\ref{bzuxy}.
At the outer periphery of the $\delta$-spots P1 and N2 away from the PIL, we find 
that the horizontal Lorentz forces are pushing the two spots together with 
a converging pattern, consistent with the converging motion as shown in 
Panel (d) of Fig.~\ref{bzuxy}. 
Meanwhile, at the PIL in the middle, apparently there is a strong repelling 
Lorentz force which pushes the two polarities in a direction away from each other. 
This repelling force naturally develops in such a system when two opposite 
polarities are pushed together with strong current sheet in between, 
while the converging motion in the $\delta$-spots is a manifestation of the 
expansion of each bipole. The combination of repelling force at the PIL and 
converging motion at the periphery gives the $\delta$-spots the observed 
curvilinear motion around one another. We here point out that this curvilinear 
motion is due to the emergence of 
the subsurface structure, which pushes and closely locks the two 
polarities P1 and N2 together during the expansion of the two bipoles. 
In \cite{vanballegooijen1989}, the presence of converging and shearing motions 
produces flux cancellation at the neutral line and the formation of a flux rope, 
which is not observed in our simulation here. The absence of such coherent a helical 
coronal flux rope may be the result of numerical diffusion
\citep{vanballegooijen2007}, given the 
large cell size of 1237.4 km in the coronal region (see Section \ref{method}). 

\begin{figure*}[ht!]
  \begin{minipage}[t] {1.0\linewidth}
    \begin{center}
      \includegraphics[width=100mm]{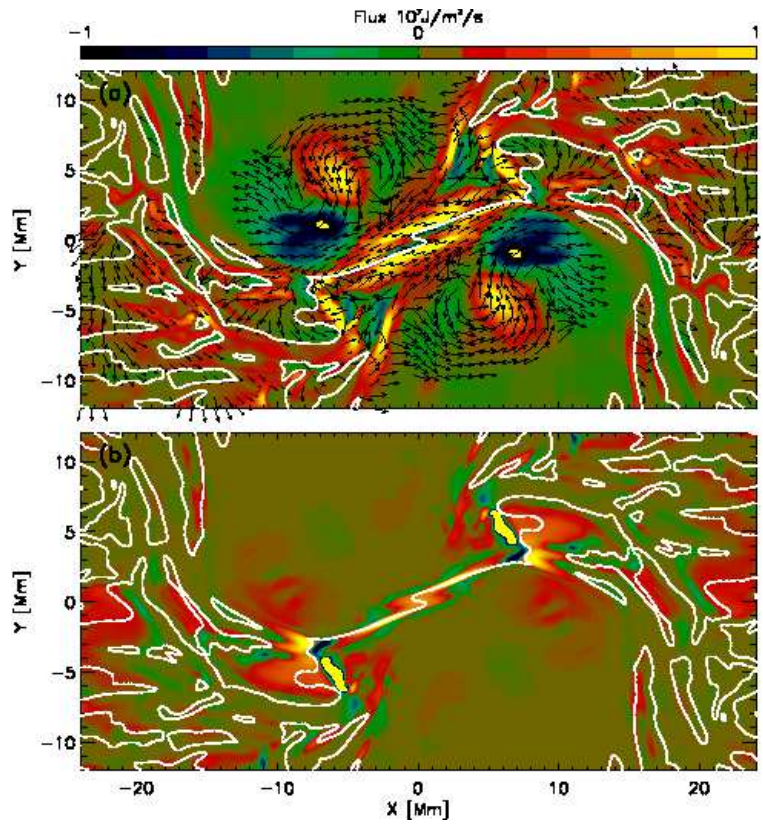}
    \end{center}
  \end{minipage}\hfill
  \caption{Poynting flux associated with the horizontal (a) and vertical (b) 
    motions of 
    the magnetic field lines at time t = 4:40:00. The black arrows represent 
    the horizontal flows in regions with $|B_z| > $ 800 G and white lines show 
    the PIL. }
  \label{poynting}
\end{figure*}

To quantify the energy transfer associated with the horizontal and vertical 
motions of the magnetic fields, we calculate the Poynting flux associated 
with the two velocity components by 
\begin{eqnarray}
  F_h& = & -\frac{1}{8\pi}\left(B_xu_{x} + B_yu_{y}\right)Bz, \\
  F_v& = &\frac{1}{8\pi} \left(B_x^2 + B_y^2\right)u_{z},
\end{eqnarray}
Fig.~\ref{poynting} shows a zoom-in view of the distribution of the energy 
fluxes associated with horizontal (a) and vertical (b) motions, 
in the central polarities (same region as Fig.~\ref{lorentz}) 
at time t = 4:40:00. 
The arrows in Panel (a) shows the horizontal flows in the $\delta$-spot 
and clearly represent a rotation pattern in each of the polarities 
as well as a strong shearing motion adjacent to the PIL shown by the white line.
Significant energy flux into the corona occurs at the PIL, identified by 
the two red bands 
in Panel (a) which coincides with the shearing motion. In addition, 
within each of the polarity, the converging motion toward the PIL shown by 
the arrows also produces a strong energy flux into the corona. 
In Panel (b), the Poynting flux by the vertical motion of the fields also 
makes a great contribution to the energy transport into the corona, with 
brightennings along the PIL, indicating strong energy flux by the 
emerging of the magnetic fields. 
The energy flux concentrated at the PIL between the two colliding spots 
transports a great amount of free energy into magnetic fields above the PIL.
As shown in Fig.~\ref{3dt0430}, the magnetic fields are highly compressed and 
non-potential above the PIL, with a free energy of $8.3\times10^{31}$ ergs,
about 46\% of the total magnetic energy of $1.8\times10^{32}$ ergs. 
This stored free energy may be the energy source of the great flares that are often
observed in $\delta$-spot regions \citep{zirin1987}. 

\section{Summary and Conclusions} \label{conclusion}

In an active-region-scale domain, 
we carry out a simulation of the evolution of a twisted magnetic flux rope 
with two buoyant sections and study the formation of a compact $\delta$-spot 
like structure during the interaction of the two emerging sections 
as well as the maintenance of 
such compact structure with a realistic size scale. 
At the beginning, the two buoyant sections rise into the atmosphere 
independently of each other, expanding dramatically in a highly stratified background 
plasma. On the photosphere, the emergence of each section forms one pair of bipoles, 
each separated from the other. 
The separation in distance decreases as the two 
sections emerge and expand. The fact that the two buoyant sections 
originate from the same flux rope prevents the two bipoles 
from escaping from each other by subsurface connection\citep{toriumi2014}.
Inevitably the leading polarity of the following bipole collides with the 
following one of the leading bipole, forming a sharp PIL with a high magnetic gradient. 
The continuous expansion during the emergence keeps pushing together the two polarities 
in the center and locks the two spots closely, producing a compact $\delta$-spot 
active region on the photosphere. 
The two colliding polarities take a curvilinear path around 
one another, producing the observed shearing and rotation motion. At the PIL, 
the horizontal velocity in the two polarities runs anti-parallel with each other. 
The magnetic fields is highly compressed and 
almost parallel to PIL in the middle of the 
$\delta$-spot. The sheared transverse magnetic field of kG strength possesses a significant 
amount of free magnetic energy up to $9.3\times10^{31}$ ergs, as a potential source for eruptions 
and flares that are often associated with $\delta$-spot active regions.
The $\delta$-spot in such an active-region-scale simulation  
produces detailed dynamics comparable with observations \citep{tang1983}, such as the 
inverted polarity against Hale's law, sharp PIL with sheared kG transverse magnetic fields,
the curvilinear motion of the polarities, as well as the complex coronal magnetic 
connectivity. 

As collision of two bipolar groups is the most popular type of formation of $\delta$-spots
\citep{zirin1987}, it is crucial to understand mechanisms that closly lock 
the two bipoles together during their emergence and expansion. 
As has been pointed out by \cite{toriumi2014}, 
the subsurface configuration plays an important role in the formation of 
a compact photospheric magnetic structure. The two emerging bipoles in our simulation, 
deeply rooted 
and connected together in the convection zone, are not able to separate from each other 
due to the internal linkage. The emergence and expansion drive the two spots into 
a curvilinear path around one another which manifests as rotation of the spots 
and shearing motion at the PIL on the photosphere. 
Nevertheless, it is the repelling force from each $\delta$-spot's bipolar 
counterpart that pushes the two polarities in the $\delta$-spot together and 
maintains the integrity of the quadupolar structure during the evolution. 
In the realistic solar convection zone, the subsurface configuration 
of multiple buoyantly emerging 
sections on the same flux rope may occur when the flux rope encounters 
multiple cells of upflowing plasma, which is possible in the turbulent interior 
\citep{nelson2011,fan2014}. 
Realistic simulations with convective motions \citep{rempel2011r,fang2012a} 
have to be implemented to study the
detailed dynamics in the emergence from the interior to the photosphere. 
In addition, in simulations with the effect of Coriolis force on the 
flux emergence, a retrograde flow naturally develops along the flux rope and 
produces assymetry in the two polarities of each emerging loop, 
with the leading one being more dominant \citep{fan1993,rempel2014}. 
Therefore, in the process where a $\delta$-spot is formed by collison of 
two bipoles, the fact that the following polarity in the $\delta$-spot is 
the leading polarity of the following bipole could also explain 
the dominance of the following polarity oftern observed in such $\delta$-spots \citep{tang1983}. 

\acknowledgments
We thank Anna Malanushenko and the referee for constructive comments and discussions. 
The work here is supported by NASA LWSCSW grant NNX13AJ04A. 
The National Center for Atmospheric Research (NCAR) is sponsored by the National 
Science Foundation.
The simulations described here were carried out on the Stampede Supercomputer 
in the Texas Advanced Computing Center (TACC) at the University of Taxas at 
Austin and and Yellowstone Supercomputer at NCAR.

\clearpage
\bibliographystyle{apj}
\bibliography{ref}


\clearpage

\end{document}